\documentclass[preprint]{aastex}
\usepackage{epsfig} 
\usepackage{amsmath}
\usepackage{amssymb}
\shorttitle{Modified Slab}
\shortauthors{F. C. Jones {\it et al.}}
\eqsecnum 
\begin{document} 
\title{The Modified Weighted Slab Technique:\\
Models and Results}

   \author{Frank C. Jones}
   \affil{Laboratory for High Energy Astrophysics, Code 660, \\
      NASA Goddard Space Flight Center, Greenbelt, MD 20771, U.S.A.}
    \email{ frank.c.jones@gsfc.nasa.gov}

   \vskip 5pt
   \author{Andrew Lukasiak}
   \affil{Institute for Physical Science and Technology,University of Maryland,\\  College Park, MD 20742, USA}
\email{al36@umail.umd.edu}

   \author{Vladimir Ptuskin}
   \affil{Institute for Terrestrial Magnetism, Ionosphere and Radio Wave Propagation of the Russian Academy of Science (IZMIRAN), Troitsk, Moscow Region 142092, Russia}
  \email{vptuskin@izmiran.troitsk.ru}

\and  
 \author{Wiliam Webber}
   \affil{New Mexico State University, Las Cruses, NM 88003, USA}
\email{}
\date{\today}

\begin{abstract} 
In an attempt to understand the source and propagation of galactic cosmic rays  we have employed the Modified Weighted Slab technique  along with recent values of the relevant cross sections to compute primary to secondary ratios including B/C and Sub-Fe/Fe for different galactic propagation models.  The models that we have considered are the disk-halo diffusion model,  the dynamical halo wind model, the turbulent diffusion model and a model with minimal reacceleration. The modified weighted slab technique will be briefly discussed and a more detailed description of the models will be given. We will also discuss the impact that the various models have on the problem of anisotropy at high energy and discuss what properties of a particular model bear on this issue.
\end{abstract}


\section{The application of the Modified Weighted Slab technique to simplified models} \label{MWSTech}
The weighted slab technique has long been used in studying the propagation of cosmic rays in the galaxy from their points of origin to their observation points near the earth \citep{Da60,GS64,GP76,Le79}, also see \citet[and  therein]{W97}. Several approximations are used in deriving this technique among them the assumption that energy loss and/or gain is not significant and that the propagation in and  loss from the galaxy may be described by a function of energy per nucleon alone. Both of these simplifications are known to be untrue, for low energies ionization energy loss can be significant and rigidity, or energy per charge, is believed to be the parameter that best describes propagation.

\citet{PJO96}  showed how the weighted slab technique could be made exact for galactic propagation models in which energy gains and losses were proportional to the same mass density that determined nuclear fragmentation and time dependent processes, eg. radioactive decay, do not play a role. This modification allows for the fact that particles had different (usually higher) energies in the past and hence different propagation properties and that propagation is considered to be a function of rigidity although energy per nucleon  is the proper parameter for nuclear fragmentation calculations. Strictly, this technique is rigorous only for models in which the particle propagation parameters are proportional to a single function of energy for each particle species, and hence does not apply to galactic wind models or turbulent diffusion models. However, most of these models may be closely approximated by simplified homogeneous models in which the mean path length has an exponential distribution  with a mean path length that is a particular function of rigidity. It is models of this type and approximation that we discuss in this study.  In this paper we present some results of the numerical simulations where the most recent set of spallation cross sections  was used. 

It should be noted that these models bear a similarity to the well known leaky box model in that they all (at least approximately) yield an exponentially decreasing path length distribution. They differ in the manner in which the mean path length varies as a function of energy. The similarity ends here, however, as the leaky box model can not predict any anisotropy as it considers the cosmic rays to be homogeneously distributed in a ``box'' of indeterminate size and shape. Furthermore, the only independent parameter in the leaky box model is the amount of matter traversed or ``grammage'' which is sufficient for stable nuclei but radioactive nuclei decay as a function of time not grammage so the more realistic models can predict significantly different results for these unstable particles. Since we will be dealing with stable nuclei only in this work our mathematics will look very much like calculations with the leaky box model but it should be remembered that we have significantly different physical models in mind when we do these calculations.

\section{Cross sections}\label{Xsections}
	The cross sections used here now include a completely updated cross
section file for the  propagation program.  This includes the new
primary cross sections in hydrogen targets for C through Ni at ~600 Mev/n
as described in \citet{Wetal98a}, as well as the hydrogen cross
sections for essentially all of the secondary nuclei from Li through Mn 
also at ~600 Mev/n reported in \citet{Wetal98c}.  The energy
dependence of these isotopic cross sections is updated and extended as well
using our earlier charge changing cross sections measured between 300 and
1700 Mev/n \citep{Wetal90} and at 15 GeV/n \citep{Wetal94}
 and assuming that the isotopic fractions are generally independent of
energy as confirmed by these earlier measurements and those of the
Transport Collaboration \citep{Ch97}.

\section{Disk-halo diffusion model.} \label{DHModel}
\begin{figure}[h]
\centering \epsfig{file=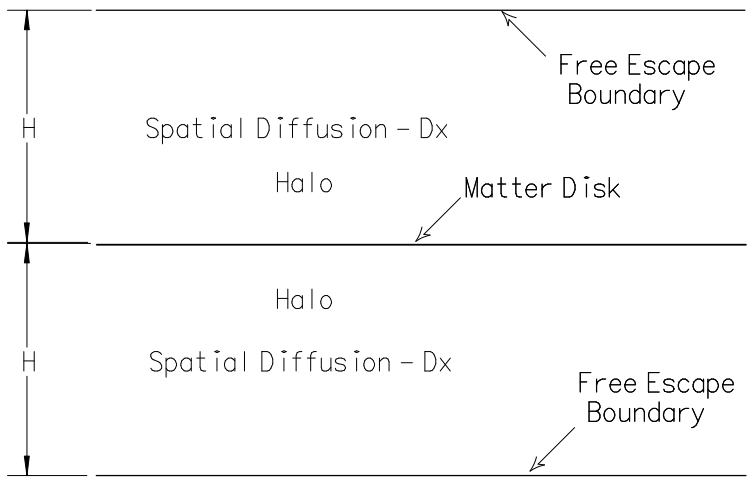,width=11cm}
\caption{Simplified galactic model with a diffusive halo,  galactic matter disk is of infinitesimal thickness.}\label{diffmod}
\end{figure}
This propagation model, \citep{GKP80,BBDGP90}schematically shown in Figure (\ref{diffmod}), involves a thin (approximately infinitesimal) disk of matter and an extended, matter free, halo through which the cosmic rays diffuse with a rigidity dependent diffusion coefficient, $D(R) $.  The distribution
function $f(z,p)$ normalized as $N=4\pi\int dpp^{2}f$ (where $N$ is the total
cosmic ray number density) obeys the equation
\begin{equation}
-\frac{\partial}{\partial z}D\frac{\partial f}{\partial z}+\frac{\mu v\sigma
}{m}\delta(z)f+\frac{1}{p^{2}}\frac{\partial}{\partial p}\left[  p^{2}\left(
\frac{dp}{dt}\right)  _{ion}f\right]  =q_{0}(p)\delta(z). \label{dif}%
\end{equation}

Here $D(p,z)$ is the particle diffusion coefficient, $\mu\approx$ 2.4
mg/cm$^{2}$ \citep{F98} is the surface mass density of the galactic disk,
$v=\beta c$ is the particle velocity, $\sigma$ is the total spallation cross
section, $\ m$ is the mean mass of interstellar atom, $\left(  dp/dt\right)
_{ion}=\mu\delta(z)b_{0}(p)/m<0$ describes the ionization energy losses and 
$q_{0}(p)\delta(z)$ is the source term that may include the yield from the
fragmentation of more heavy nuclei.

We shall assume that diffusion does not depend on position, i.e. $D=D(p)$.
There is a cosmic ray halo boundary at $\shortmid z\shortmid=H$ where cosmic
rays freely exit from the Galaxy.

Integrating equation (\ref{dif}) in the vicinity of galactic plane $\lim
\int_{-\varepsilon}^{\varepsilon}dz(...)$ at $\varepsilon\longrightarrow0 $
one can find the boundary condition at $z=0+\varepsilon:$%

\begin{equation}
-2D\frac{\partial f_{0}}{\partial z}+\frac{\mu v\sigma}{m}f_{0}+\frac{1}%
{p^{2}}\frac{\partial}{\partial p}\left[  p^{2}\frac{\mu b_{0}}{m}%
f_{0}\right]  =q_{0}(p), \label{plane}%
\end{equation}
where $f_{0}(p)=f(z=0,p)$ is the distribution function at the galactic midplane.

The solution of equation (\ref{dif}) at $z\neq0$ under the boundary condition
$f(\shortmid z\shortmid=H)=0$ is:
\begin{equation}
f=f_{0}\frac{H-\shortmid z\shortmid}{H}. \label{halo}%
\end{equation}

Calculating $-D\frac{\partial f_{0}}{\partial z}$ from equation (\ref{halo})
and substituting it in equation (\ref{plane}), one can get the following
closed equation for $f_{0}(p):$%
\begin{equation}
\frac{2Df_{0}}{\mu vH}+\frac{\sigma}{m}f_{0}+\frac{1}{p^{2}}\frac{\partial
}{\partial p}\left[  p^{2}\frac{b_{0}}{mv}f_{0}\right]  =\frac{q_{0}}{\mu v}.
\label{closed-plane}%
\end{equation}

Now we introduce cosmic ray intensity as function of kinetic energy per
nucleon $I(E_{k})dE_{k}=vf_{0}(p)p^{2}dp,$ so that $f_{0}(p)=I(E_{k})A/p^{2}$
($A$ is the atomic number). Equation (\ref{closed-plane}) gives the following
equation for $I(E_{k}):$%
\begin{equation}
\frac{I}{X_{dif}}+\frac{d}{dE_{k}}\left[  \left(  \frac{dE_{k}}{dx}\right)
_{ion}I\right]  +\frac{\sigma}{m}I=Q \label{Lbox}%
\end{equation}
with the effective escape length for diffusing particles
\begin{equation}
X_{dif}=\frac{\mu vH}{2D}, \label{Xdif_2}%
\end{equation}
and the source term%

\begin{equation}
Q=\frac{q_{0}(p)p^{2}A}{\mu v}. \label{source}%
\end{equation}.

Equation (\ref{Lbox}) is identical to the transport equation for cosmic rays
in the so called leaky box model, the popular empirical model used in the
studies of cosmic ray propagation. The leaky box model has the only parameter,
the escape length $X_{lb}$, which describes the cosmic ray propagation and
this parameter is considered as an empirical characteristic determined from
cosmic ray data. The physical meaning of the escape length is that it is equal
to the mean thickness of matter traversed by cosmic rays before exit from the Galaxy.

The foregoing consideration confirms the well known result that the diffusion
model with relatively thin galactic disk (with half thickness $h\ll H$) and
with flat halo (radius of the galactic disk $R\gg H$) is almost equivalent to
the leaky box model for calculation of abundance of stable nuclei in cosmic
ray \citep{BBDGP90, PJO&S97}. The
equivalence holds for not very heavy nuclei which have the total cross
sections $\sigma\ll\frac{m}{X_{dif}}\times\frac{H}{h}$. Under this condition,
the relation between the parameters of the diffusion model and the equivalent
leaky model follows from the equation $X_{dif}=X_{lb}$ that leads to the
expression for diffusion coefficient%

\begin{equation}
D=\mu\beta cH/(2X_{lb}), \label{Diffcoef}%
\end{equation}
where $\beta=v/c$.
In this simplified model the path length distribution is a decreasing exponential function with a mean path length dependent on energy. Such a dependence is shown in equation (\ref{grammage})) where the parameters $X_0$, $R_0$, and $a$ are determined by the density of the matter disk, the size of the halo and the spectrum of magnetic turbulence that sets the diffusion coefficient's rigidity dependence. 
\begin{equation}
X_{lb}=X_0\beta\text{ g/cm}^{2}\text{ at }R<R_0\text{GV},\text{ }
X_{lb}=X_0\beta(R/R_0\text{GV})^{-a}\text{g/cm}^{\text{2}}\text{ at }
R\geq R_0 \text{GV}\label{grammage}
\end{equation}
where $R$ is the particle rigidity and $\beta = v/c$.
This parameterization is valid for particles in the interstellar medium with
energies from about 0.4 GeV/n to 300 GeV/n where data on secondary nuclei
are available.  It is worth noting that the same escape length (equation (\ref{grammage}))
satisfactorily reproduces both B/C and Sub-Fe/Fe ratios (no need for path
length ''truncation''). It should be borne in mind, however, that this diffusion model does not predict the sudden change of the mean grammage as a function of energy that is displayed in equation (\ref{grammage}) at a rigidity of $R_0$ GV or any other rigidity. This is a strictly (though widely accepted) {\it ad hoc} construction that appears to be required by the data. We will see that the  models following this one do not require this behavior to be externally imposed.

The physical interpretation of the empirical equation (\ref{grammage}) for $R> R_0$ GV {\em can} be given in the framework of the diffusion model by referring to equation (\ref{Diffcoef}).
The theory of particle resonant scattering and diffusion in the turbulent
interstellar medium predicts the scaling of the diffusion coefficient
$D_{res}=\kappa\beta R^{a}$, where constant $\kappa$ is determined by the
level of hydromagnetic turbulence with the spectrum $W_{k}dk\propto k^{-2+a}dk,$
$a=$const. The scattering of particles with Larmor radius $r_{g}$ is mainly
due to the interaction with inhomogeneities of the scale $1/k\approx r_{g}$. Since the time to diffuse to the galactic boundary, and hence the mean grammage traversed, is inversely proportional to the diffusion coefficient the interpretation is clear.
The observations of interstellar turbulence are consistent with the existence of a single power-law spectrum with $0.2 \leq a \leq 0.6$ at wave numbers $10^{-20} \text{cm}^{-1} \leq k \leq 10^{-8}\text{cm}^{-1} $, see \citet{Retal88}. The Kolmogorov spectrum, usually considered to be representative of the interstellar spectrum corresponds to $a=1/3$.

\section{Galactic wind model.}\label{GWmodel}
\begin{figure}[h]
\centering \epsfig{file=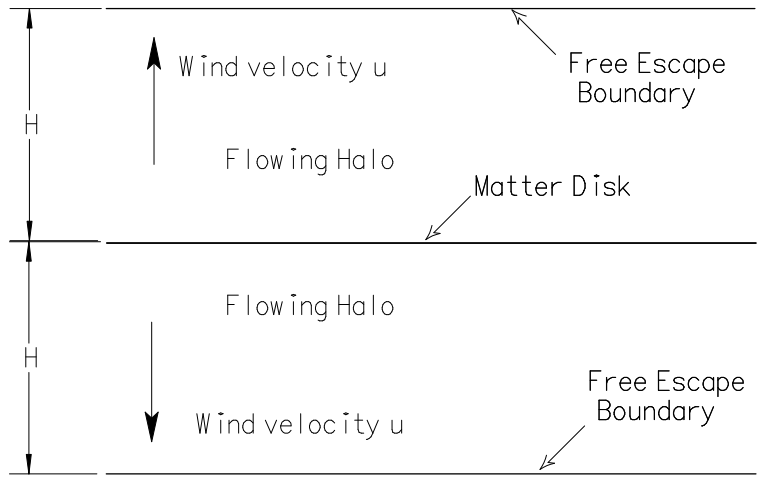,width=11cm}
\caption{Simplified galactic model with a halo wind; galactic matter disk is of infinitesimal thickness.}\label{Wind_model}
\end{figure}

We consider a one-dimensional galaxy model shown in Figure (\ref{Wind_model}) (the coordinate $z$ is perpendicular to
the galactic plane). Cosmic ray sources and interstellar gas are concentrated
in a thin disk at $z=0$ (For details of this model see \citet{Jok76, Jon79}). The distribution function $f(z,p)$ obeys
the equation
\begin{equation}
-\frac{\partial}{\partial z}D\frac{\partial f}{\partial z}+u\frac{\partial
f}{\partial z}-\frac{du}{dz}\frac{p}{3}\frac{\partial f}{\partial p}+\frac{\mu
v\sigma}{m}\delta(z)f+\frac{1}{p^{2}}\frac{\partial}{\partial p}\left[
p^{2}\left(  \frac{dp}{dt}\right)  _{ion}f\right]  =q_{0}(p)\delta(z).\label{dif-con}
\end{equation}
where $u$ is the wind convection velocity.

We shall assume that 
the wind velocity $u$ is constant and directed outward from the galactic plane.
There is a cosmic ray halo boundary at $\shortmid z\shortmid=H$ where cosmic
rays freely exit from the Galaxy.
Using the same prodecure as in section (\ref{DHModel}) we integrate equation (\ref{dif-con}) in the vicinity of galactic plane ($\lim_{\varepsilon\rightarrow0}{\int_{-\varepsilon
}^{\varepsilon}}dz(...)$) one can find the
boundary condition at $z=0+\varepsilon:$

\begin{equation}
-2D\frac{\partial f_{0}}{\partial z}-\frac{2up}{3}\frac{\partial f_{0}}{\partial p}+\frac{\mu v\sigma}{m}f_{0}
+\frac{1}{p^{2}}\frac{\partial}{\partial p}\left[  p^{2}\frac{\mu b_{0}}{m}f_{0}\right]  =q_{0}(p),\label{in-plane}
\end{equation}

where $f_{0}(p)=f(z=0,p)$ is the distribution function at the galactic midplane.

The solution of equation (\ref{dif-con}) at $z\neq0$ under the boundary condition $f(\shortmid
z\shortmid=H)=0$ is:
\begin{equation}
f=f_{0}\frac{1-\exp\left(  -u(H-\shortmid z\shortmid)/D\right)  }%
{1-\exp\left(  -uH/D\right)  }.\label{halo-soln}
\end{equation}

Calculating $-D\frac{\partial f_{0}}{\partial z}$ from equation (\ref{halo-soln}) and substituting it
in equation (\ref{in-plane}), one can get the following  equation for $f_{0}(p):$%
\begin{equation}
\frac{2uf_{0}}{\mu v\left(  \exp(uH/D)-1\right)  }-\frac{2u}{3\mu v}%
p\frac{\partial f_{0}}{\partial p}+\frac{\sigma}{m}f_{0}+\frac{1}{p^{2}}%
\frac{\partial}{\partial p}\left[  p^{2}\frac{b_{0}}{mv}f_{0}\right]
=\frac{q_{0}}{\mu v}. \label{closed-in-plane}
\end{equation}

Equation (\ref{closed-in-plane}) gives the following equation for the cosmic-ray intensity 
$I(E_{k}):$%
\begin{equation}
\frac{I}{X_{w}}+\frac{d}{dE_{k}}\left[  \left(  \left(  \frac{dE_{k}}%
{dx}\right)  _{ad}+\left(  \frac{dE_{k}}{dx}\right)  _{ion}\right)  I\right]
+\frac{\sigma}{m}I=\frac{q_{0}(p)p^{2}A}{\mu v} \label{Lbox_2}
\end{equation}
with the effective escape length
\begin{equation}
X_{w}=\frac{\mu v}{2u}\left(  1-\exp\left(  -uH/D\right)  \right)  , \label{X}
\end{equation}
and the adiabatic energy loss rate per g/cm$^{2}$%
\begin{equation}
\left(  \frac{dE_{k}}{dx}\right)  _{ad}=-\frac{2u}{3\mu c}\sqrt{E_{k}%
(E_{k}+2E_{0})}.\label{adloss}
\end{equation}
At this point  we have arrived at an equation that is essentially the same as for the thin disk-halo  model in the sense that the path length distribution will be an exponential with the mean path length given by $X_w$. We may therefore apply the modified weighted slab technique to solve the equations.

As before, the asymptotic
scaling of $X_{w}$ in this case is
\begin{equation}
X_{w}=\frac{\mu v}{2u}\propto v\text{ at small rigidities (when }uH/D\gg1),\label{xatsmall}
\end{equation}
and
\begin{equation}
X_{w}=\frac{\mu vH}{2D}\propto R^{-a}\text{ at large rigidities (when }uH/D\ll1). \label{xatlarge}.
\end{equation}
It should be noted that here the flattening of the curve of mean grammage vs. energy at low energy arises naturally from the model and does not need to be inserted by hand.

Equation (\ref{X}) may be presented in the following form useful to fit the
observations:
\begin{equation}
X_{w}=\beta X_{0}\left(  1-\exp\left(  -\frac{1}{\beta\left(  R/R_{0}\right)
^{a}}\right)  \right)  .
\end{equation}Here $X_{0}=(\mu c)/(2u),$ and $R_{0}=\left(  uH/\kappa_{0}\right)  ^{1/a}.$ 
The adiabatic energy loss term, equation (\ref{adloss}) in these notations reads as
\begin{equation}
\left(  \frac{dE_{k}}{dx}\right)  _{ad}=-\frac{1}{3X_{0}}\sqrt{E_{k}%
(E_{k}+2E_{0})}.
\end{equation}

It is worth noting that here we consider only a simple model with constant wind velocity.  In more realistic models the wind velocity depends on th distance from the galactic plane, {\it e.g.} in  a self consistent model of a cosmic ray driven galactic wind \citep{PVZB97}.
\section{Turbulent diffusion}\label{Tdiff}

\begin{figure}[h]
\centering \epsfig{file=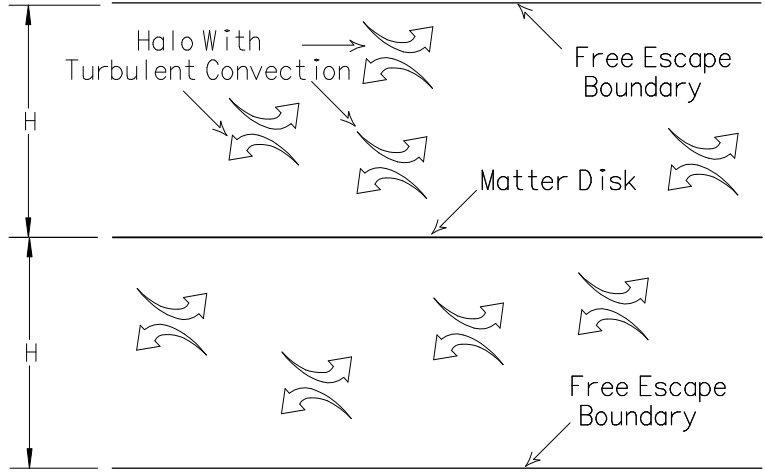,width=11cm}
\caption{Turbulent diffusion model, halo is diffusive with two types of diffusion.}\label{turb-model}
\end{figure}

In this model there is no regular convective (wind) transport, rather the flow is turbulent and the transport of the particles is more random. In fact the convection in this flow is of the same nature as diffusion with the diffusive properties determined by the random flow.  (See Figure (\ref{turb-model})) Equation (\ref{dif-con})
becomes

\begin{equation}
-\frac{\partial}{\partial z}D\frac{\partial f}{\partial z}+\frac{\mu v\sigma
}{m}\delta(z)f+\frac{1}{p^{2}}\frac{\partial}{\partial p}\left[  p^{2}\left(
\frac{dp}{dt}\right)  _{ion}f\right]  =q_{0}(p)\delta(z). \label{turb}
\end{equation}.

The equation for $I(E_{k})$ is the following%

\begin{equation}
\frac{I}{X_{dif}}+\frac{d}{dE_{k}}\left[  \left(  \frac{dE_{k}}{dx}\right)
_{ion}I\right]  +\frac{\sigma}{m}I=\frac{q_{0}(p)p^{2}A}{\mu v} \label{turb-LB}
\end{equation}
where
\begin{equation}
X_{dif}=\frac{\mu vH}{2D} \label{X-turb}
\end{equation}.
We assume now that cosmic ray diffusion is provided simultaneously by
turbulent diffusion with the diffusion coefficient $D_{t}$ that does not
depend on particle energy (may be estimated as $D_{t}=u_{t}L_{t}/3$, where
$u_{t}$ and $L_{t}$ are the characteristic random velocity and correlation
scale of large-scale turbulent motions of the interstellar gas) and by
resonant diffusion with the diffusion coefficient $D_{res}=\beta\kappa
_{0}R^{a}$ (here $\kappa_{0}=const$, $a=const$)\ provided by the scattering on
hydromagnetic turbulence as was discussed in the previous section. The total
diffusion coefficient which appears in equation (\ref{X-turb}) is equal to
\begin{equation}
D=D_{t}+D_{res}. \label{D-comb}
\end{equation}.

Equation (\ref{X-turb}) may be presented in the following form that is useful to fit the
observations:
\begin{equation}
X_{dif}=\frac{\beta X_{0}}{1+\beta\left(  R/R_{0}\right)  ^{a}}. \label{Xdif}
\end{equation}
Here $X_{0}=(\mu cH)/(2D_{t}),$ and $R_{0}=(D_{t}/\kappa_{0})^{1/a}.$
 
We note again the flattening of the mean grammage vs. energy curve for low rigidities, as in the previous wind model, arises naturally from the physical picture.  Actually the reason is the same in both models; below a particular rigidity kinetic diffusion becomes slower than the convective transport in removing particles from the galaxy and the particles residence time becomes independent of rigidity or energy. 

\section{Stochastic Reacceleration.}\label{StR}

\begin{figure}[h]
\centering \epsfig{file=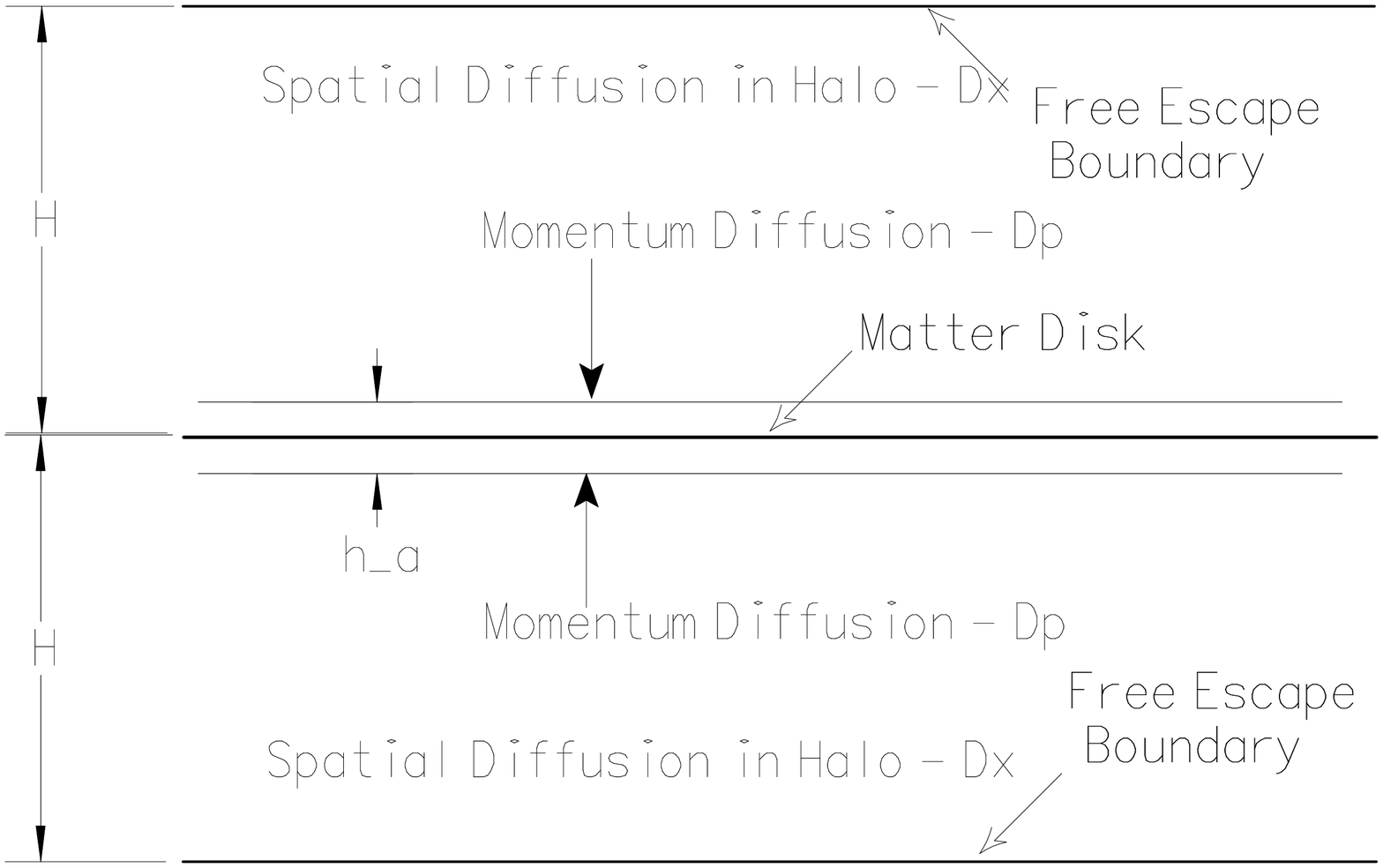,width=11cm}
\caption{ Reacceleration model, halo is diffusive in momentum as well as space.}\label{stoch-model}
\end{figure}

This model  \citep{SP94} assumes no convective or wind motion in the system( see Figure (\ref{stoch-model})). As before, the spatial
diffusion is provided by scattering on random hydromagnetic waves with a
 spectrum that results into $D\propto vR^{a}$.   For a Kolmogorov spectrum of turbulence $a=1/3$;
but we shall consider the value of $a$ as a free
parameter, in the same way as in the models
discussed above. This determines the behavior of the escape length $X_{e}
=X_{0}R^{-a}$\ (see equation (\ref{xatlarge})) at all energies. So far this does not differ from the diffusion model but here we consider the fact that the turbulence is not static, rather, the magnetic fluctuations move with the Alfv\'en velocity producing a diffusion in momentum as well as in space. Stochastic acceleration
with a diffusion coefficient in momentum $K\thicksim p^{2}v_{A}^{2}/D$ ($v_{A}$
is the Alfv\'en velocity) essentially modifies  the spectra of primaries and
secondaries below about $10$ GeV/n and can produce the characteristic peak in
$B/C$ ratio at few GV. The acceleration becomes inefficient at high energies
and the model is reduced to a simple diffusion model without reacceleration and with
the escape length $X_{e}=X_{0}R^{-a}$\ \ \ \ \ at $\ E>20-30$ GeV/n.

We use the same notations as Seo and Ptuskin and use their equation (12) for
the calculations:
\begin{align}
&\frac{I}{X_{e}}+\frac{\sigma}{m}I+\frac{d}{dE}\left[  \left(  \frac{dE}
{dx}\right)  _{ion}I\right]   +\alpha\left\{  \left[  \frac{A}{Z}\frac{(E_{k}+E_{0})}{\beta}\frac{dX_{e}
}{dR}+X_{e}\right]  I\right.\\
&-\left.\frac{A}{2Z}\beta(E_{k}+E_{0})^{2}\frac{dX_{e}}{dR}
\frac{dI}{dE}-\frac{\beta^{2}}{2}(E_{k}+E_{0})^{2}X_{e}\frac{d^{2}I}{dE^{2}
}\right\} \nonumber\\
&  =\frac{q_{0}(p)p^{2}A}{\mu v}.\nonumber
\label{stoch-LB}
\end{align}
Here the parameter $\alpha$, which is defined as
\begin{equation}
\alpha=\frac{32}{3a(4-a^{2})(4-a)}\frac{h_{a}}{H}\left(  \frac{v_{a}}{\mu
c}\right)  ^{2} \label{alpha}
\end{equation}
(its dimension is (g/cm$^{2})^{-2})$, determines the efficiency of
reacceleration, $h_{a}$ is the height of reacceleration region. In this work we have taken $h_a/H=1/3$.        

The parameters we have to find from fitting the data are $X_{0}$,  $\alpha$
and $a$ since we will not prescribe the spectrum of the interstellar turbulence.

\section{Fits to data}\label{Fits}
We fit the four models discussed above to a collection of data compiled by  \citet{SS98} for the B/C and for the Sub-Fe/Fe ratios. We determined which parameters  best fit (in the weighted least squares sense) {\em both ratios simultaneously}  for each model. The parameters found for each model are given in Table (\ref{params}). It should be noted that because considerable computing is required for each set of parameter values the search in parameter space was not automated. It was performed by hand and thus we can not guarantee that the fits that we have found are rigorously {\em least} squares fits, they are simply the smallest values we could find in our searches. 
\begin{table}[h]
\begin{center}
\begin{tabular}[h]{r|c|c|c|c}
\multicolumn{5}{c}{Fitted Parameters}\\
\hline
Model & $X_0$ (g/cm$^2$) & $R_0$  (GV)& a & $\chi^2$ (normalised)\\
\hline
Disk-Halo Diffusion& 11.8&4.9&0.54&1.3\\
Wind&12.5&11.8&0.74&1.5\\
Turbulent Diffusion& 14.5&15.0&0.85&1.8\\
Stochastic Reacceleration& 9.4&$\alpha = 2.6\times 10^{-3}$&0.30&1.8\end{tabular}
\caption{Best fit parameters for fitting secondary/primary ratios for the models considered. For Minimal Reacceleration there is no $R_0$ parameter, rather the strength of acceleration is given by the dimensionless parameter $\alpha$. \label{params}}
\end{center}
\end{table}

These parameters imply physical quantities for the different models given in Table (\ref{PhysicalParams}).  Notice that certain parameters are meaningful for specific models only, not all models in general.
\begin{table}[h]
\begin{center}
\begin{tabular}[h]{r|c|c|c|c}
\multicolumn{5}{c}{Physical Parameters}\\
\hline
Model & $D_{res}$ (cm$^2$/s) & $D_t$  (cm$^2$/s)&$u$ (km/s)& $V_a$ (km/s)\\
\hline
Disk-Halo Diffusion&$2.0\times 10^{28}\beta$&-&-&-\\
Wind&$7.2\times 10^{27}\beta$&-&29&-\\
Turbulent Diffusion& $3.8\times 10^{27}\beta$&$3.8\times 10^{28}$&-&-\\
Stochastic Reacceleration& $5.9\times 10^{28}\beta$&-&-&40
\end{tabular}
\caption{Physical parameters implied by the fit parameters of Table \protect\ref{params}. The values of $D_{res}$ should be multiplied by $R^a$ with $R$ in GV and $a$ taken from Table \protect\ref{params}}. \label{PhysicalParams}
\end{center}
\end{table}

In the turbulent diffusion model the turbulent diffusion coefficient can be estimated as $D_{t}=u_{t}L/3$, where
$u_{t}$ is the characteristic velocity and $L$ is the characteristic scale of
the turbulent motions. Since we would not expect the magnitude of the velocity $u_{t}$ to exceed 100
km/s that leads to the very large value of $L\geq0.76H$. Thus this
model of cosmic ray  transport requires a value of $D_{t}$ that is difficult
to reconcile with acceptable values of parameters of the interstellar turbulence. 

\begin{figure}[h]
\centering \epsfig{file=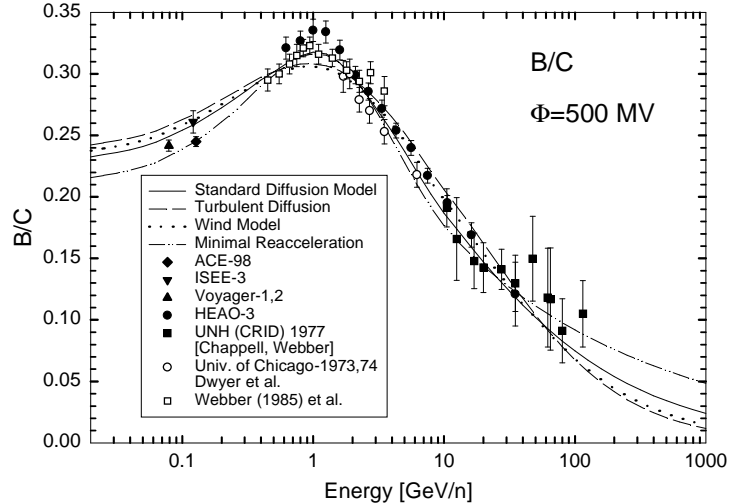,width=11cm}
\caption{  The least squares fit to observed B/C ratios.
in four propagation models: turbulent diffusion (dashed lines), wind (dotted
lines), reacceleration (dash-dotted lines), and the disk-halo diffusion model
with the diffusion coefficient given by equation (\ref{Diffcoef}) (solid lines). $\Phi$ is the force field approximation solar modulation parameter. Data are from a
comprehensive compilation by \citet{SS98}. 
}\label{B2C}
\end{figure}
\begin{figure}[h]
\centering \epsfig{file=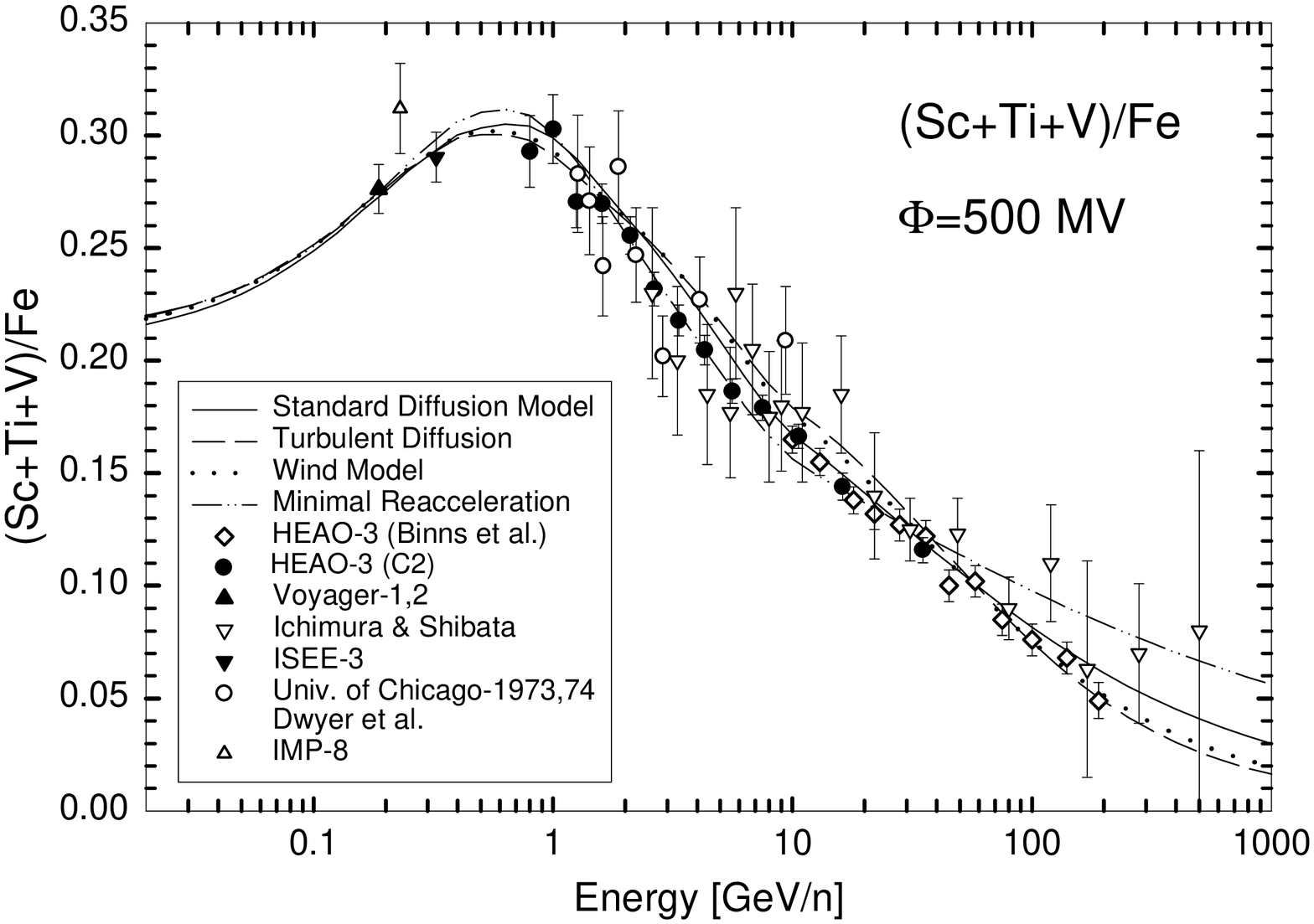,width=11cm}
\caption{  The least squares fit to observed  Sub-Fe/Fe ratios.
in four propagation models: turbulent diffusion (dashed lines), wind (dotted
lines), reacceleration (dash-dotted lines), and the disk-halo diffusion model
with the diffusion coefficient given by equation (\ref{Diffcoef}) (solid lines).  $\Phi$ is the force field approximation solar modulation parameter. Data are from a
comprehensive compilation by \citet{SS98}. 
}\label{sFe2Fe}
\end{figure}

These fits are displayed with the data in Figures (\ref{B2C}) and (\ref{sFe2Fe}). After determining the best fit parameters we then used them to propagate  primary spectra for C, O and Fe as discussed in
Section (\ref{source_section}).  If one compares Figures (\ref{B2C}) and (\ref{sFe2Fe}) with Figures (\ref{Carbon}) and (\ref{Iron}) in Section (\ref{source_section}) one can see how much more sensitive secondary to primary ratios are to the propagation model than are simple primary spectra.  Although the models were all chosen to fit the same secondary to primary data one can still see significant differences in the model curves whereas in the plots of carbon and iron the model curves are quite hard to distinguish from one another.

\section{Source spectrum.}
\label{source_section}
\begin{figure}[h]
\centering \epsfig{file=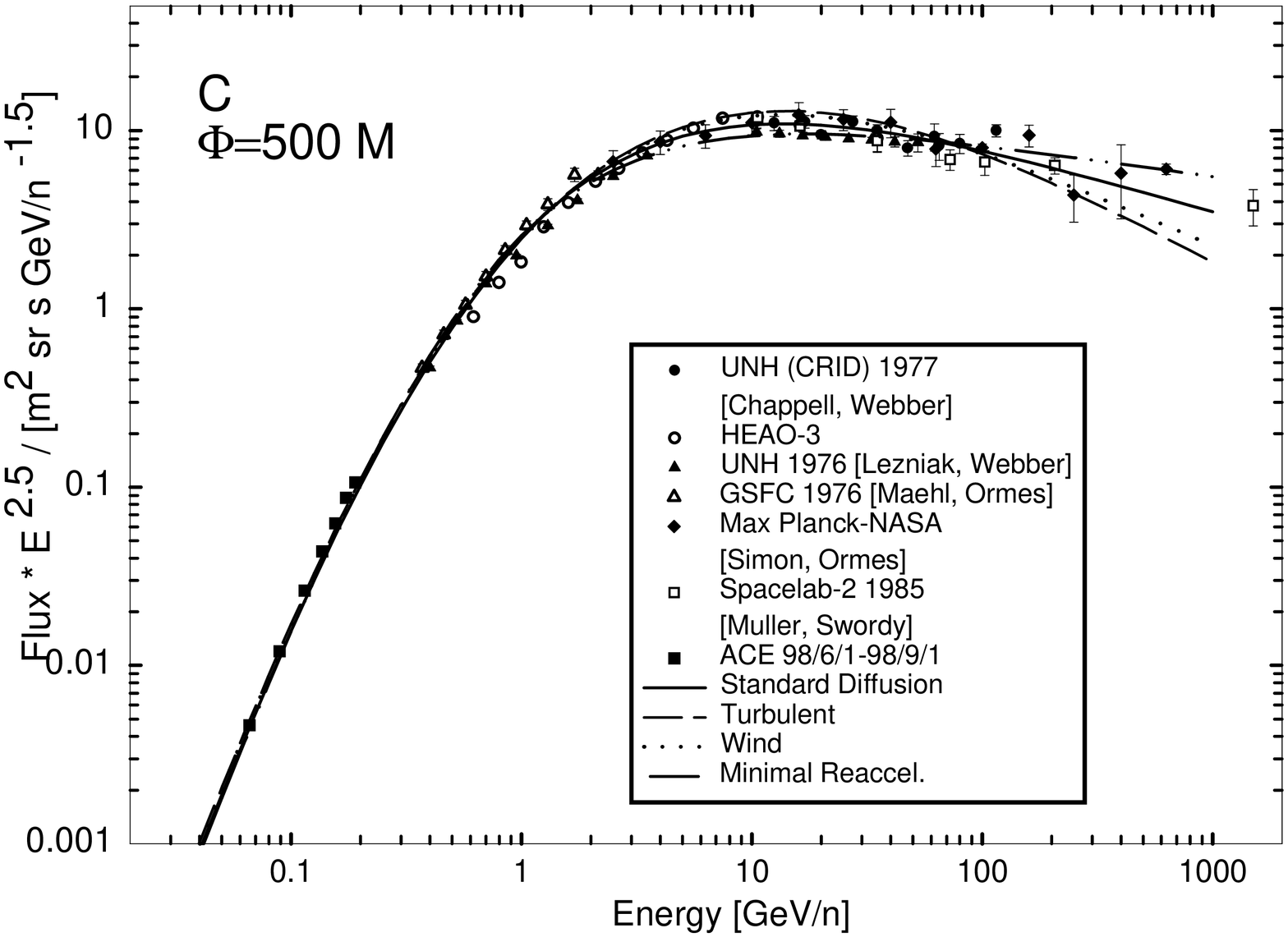,width=11cm}
\caption{ The least squares fit to the C spectrum in the same
propagation models as indicated in Figures (\ref{B2C}) and (\ref{sFe2Fe}). }\label{Carbon}
\end{figure}

\begin{figure}[h]
\centering \epsfig{file=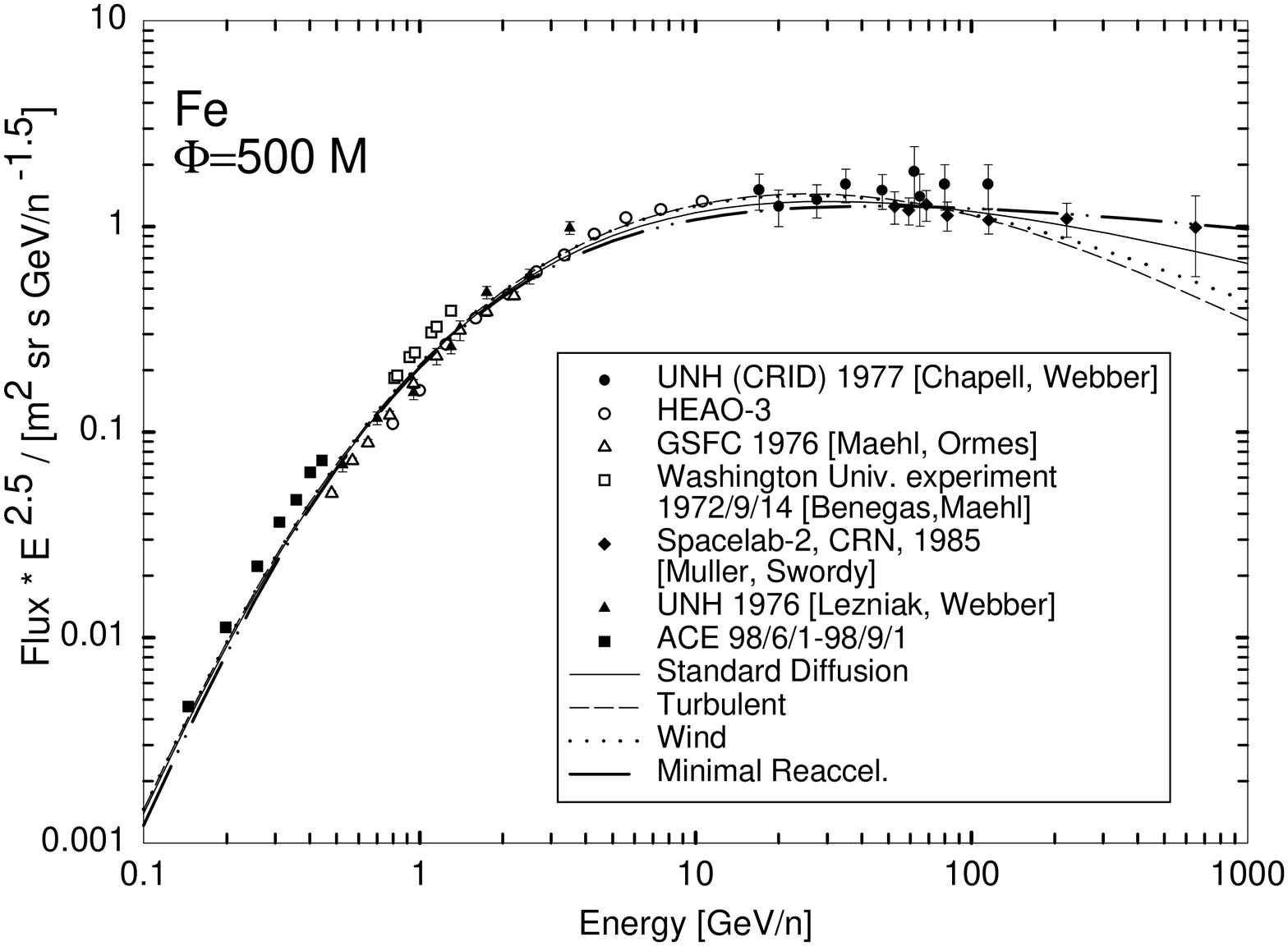,width=11cm}
\caption{ The least squares fit to the Fe spectrum in the same
propagation models as indicated in Figures (\ref{B2C}) and (\ref{sFe2Fe}). }\label{Iron}
\end{figure}

Using the above determined propagation models, we tested different source
spectra to fit observed spectra of primary nuclei C, Fe at energies
0.5-100 GeV/n. The best fit for the disk-halo diffusion, wind and turbulent diffusion model is provided by the source spectrum $Q\propto
R^{-2.35}$;  for the minimal reacceleration model the best fit was obtained with  $Q\propto \beta R^{-2.40}$ Figures (\ref{Carbon}) and (\ref{Iron}) show how the different propagation models fit the observations assuming these source spectra. One can expect that asymptotically at very high energies, $E>$ 100 GeV/n, the
relation $\gamma=\gamma_{s}+a$, where $\gamma(\gamma_s)$ is the exponent of the observed (source) 
differential spectrum $I(E)\propto E^{-\gamma}$) is fulfilled. It is of interest that three of the four models were well fit with a source spectrum that was a simple power law in rigidity. Only the minimal reacceleration model required a modification at low (non-relativistic) energies. A similar behavior was found by \citet{HS95} in their investigation of a reacceleration model.

\section{Anisotropy constraint}\label{AnCon}

The observed galactic cosmic rays are highly isotropic. The amplitude of the first
angular harmonic of the cosmic ray distribution in the interstellar medium near
the solar system is approximately equal to $\delta=(0.5-1.0)\times10^{-3}$ at
energies 10$^{12}-10^{14}$ eV \citep{NFS89,CG91,AAA93,AAC93}. The anisotropy is almost
energy-independent but a slow increase with energy or even irregular behavior at
energies close to 10$^{14}$ eV are not excluded.

Possible interpretation of these observations is based on a contribution of
local cosmic ray sources \citep{DGP91}. Supernovae and their remnants
are assumed to be the instantaneous point sources. The cosmic ray density is
determined by the total contribution from numerous sources (SN outbursts occur
in about each 30 yr, whereas the cosmic ray confinement time in the Galaxy
exceeds 10$^{7}$ yr). However, the anisotropy may be defined by an individual
nearby source. If the diffusion coefficient $D(E)$ increases with energy, the
contribution of an individual outburst at distance $r$ gives rise to an
anisotropy which depends nonmonotonicly on energy. A source's contribution to the  anisotropy amplitude
reaches its maximum when $t\sim r^{2}/D $, where $t$ is the age of
instantaneous source. Analysis of the list of supernova remnants and
pulsars indicates that Geminga, Vela, Lupus Loop, Loop III, and some others
may prove to be the sources sustaining the anisotropy observed at 10$^{12}%
$-10$^{14}$ eV.

One has to check the compatibility of the propagation models discussed in the
previous sections with the data on cosmic ray anisotropy. Each of these models
assumes a specific energy dependence of cosmic ray leakage from the Galaxy. It
might result in an anisotropy which exceeds the observational limit. Effect of
the solar wind masks the anisotropy of galactic comic rays at energies less
than about 10$^{12}$ eV for an observer at the Earth. Below we assume that the
dependence of cosmic ray diffusion on energy determined from the observations
of secondary to primary ratios in cosmic rays at energies up to about
10$^{11}$ eV can be extrapolated to energies 10$^{12}-10^{14}$ eV. This
simplifying assumption is based on the observation that there is no any
drastic change in the total cosmic ray energy spectrum in the energy range
10$^{10}$ - 10$^{14}$ eV that would be indicative of change in energy
dependence of cosmic ray transport.

The equation for the amplitude of cosmic ray anisotropy perpendicular to
galactic plane is the following (see {\it e.g.~}\citet{BBDGP90}):
\begin{equation}
\delta=-\frac{3}{\text{v}f}\left(  D\frac{\partial f}{\partial z}+u\frac{p}{3}\frac{\partial f}{\partial p}\right)  \label{anisotropy}.
\end{equation}
which includes both the diffusion and convection fluxes of cosmic rays.

It is easy to show that diffusion dominates over convection at high enough
energies $E>100$ GeV and that effects of ionization energy losses and
reacceleration are not essential at these energies in the models we discuss
here. Notice also that cosmic ray anisotropy is mainly determined by the most
abundant proton component of cosmic rays that is subject to insignificant
effect of nuclear interaction with the interstellar gas. In these conditions,
the second term in brackets in the last equation can be omitted, and the
cosmic ray transport equation can be presented in a simple form as
\begin{equation}
-\frac{\partial}{\partial z}D\frac{\partial f}{\partial z}=q_{0}(p)\delta(z).
\end{equation}
Now it is easy to show that the anisotropy for an observer just above the
$\delta$-plane with cosmic ray sources is approximately equal to
\begin{equation}
\delta_{0}\approx\frac{3q_{0}(p)}{2\text{v}f_{0}}\approx\frac{3\mu}{2X}%
\end{equation}
in all diffusion models . Here $X$ is the escape length which obeys the
equation $X\approx\mu\beta cH/(2D)$ at high particle energies.

Cosmic ray anisotropy inside the source region is smaller than $\delta_{0}.$
For cosmic ray sources uniformly distributed through the disk with a total
thickness $2h$, the anisotropy at distance $z$ from the central galactic plane
is estimated as $\delta_{z}\approx\delta_{0}z/h$ at $\left|  z\right|  <h$ \citep{P97}
. With parameters found in the previous sections and
extrapolated to energy $10^{14}$ eV and with assumption that $z/h=0.1$ ($z=20$
pc, $h=200$ pc), we have the following expected values of the anisotropy
$\delta_{z}(10^{14}$eV$)$\ which is due to the streaming of cosmic rays
perpendicular to galactic plane: $7\times10^{-3}$ in the basic diffusion
model, $4\times10^{-2}$ in the model with turbulent diffusion, $2\times
10^{-2}$ in the galactic wind model, $1\times10^{-3}$ in the minimal
reacceleration model. These values are uncomfortably large, even for the minimal reacceleration model

Of course, the models of the galaxy that we have employed here are highly simplified, primarily in the high degree of symmetry and smoothness that they exhibit.  This means that the anisotropies that we have calculated  are {\em most likely lower limits} to those that would be obtained from more complex models. It is always possible to construct models with the solar system in a rather favored position of symmetry that would produce a smaller value of the anisotropy but such models, lacking any particular knowledge of their reality, are {\it a priori} unlikely.

\section{Conclusion.}\label{Concl}
The objective of the present work was to investigate several  models of cosmic ray propagation
and nuclear fragmentation in the interstellar medium using the most
recent set of spallation cross sections and employing  the
modified weighted slab method which allows us to obtain  exact solutions. 
It is clear from a comparison of Figures (\ref{B2C}) and (\ref{sFe2Fe}) on the one hand and Figures (\ref{Carbon}) and (\ref{Iron}) on the other
that the secondary to primary ratio is a much more sensitive function of the propagation model (and relatively insensitive to the assumed source spectrum) than is the  relation of an observed spectrum to the source spectrum. It is for just this reason that we used this ratio to test our various models before then deducing primary spectra.
The standard flat-halo diffusion model, the models with turbulent
diffusion, the model with constant wind velocity, and the diffusion model with reacceleration were tested. For each model we then picked the parameter set that best fit the data. The primary spectra were then calculated for each model using these parameters. 
All these models are able to explain the decrease of secondary/primary ratios at
energies below a few GeV/n. The turbulent diffusion and the wind transport
work simultaneously with resonant diffusion. The last process dominates in
cosmic ray transport at high energies. In order to reproduce a sufficiently
sharp bend in the secondary/primary ratio in these models the diffusion must have a strong dependence on rigidity
$D_{res}\propto\beta R^{a}$, where $a=$ 0.74 - 0.85, at least at low energy ($E< 30$ GeV/n). However, this could lead to a problem. The extrapolation of such strong rigidity dependence of diffusion to
energies $10^{3}-10^{5}$ GeV produces anisotropies that are at strong variance
with the observed anisotropy at these energies.

The disk-halo diffusion model has less severe problem with the explanation of
observed small anisotropy than the models with turbulent diffusion and wind.
Also, the scaling of diffusion coefficient on rigidity $D_{res}\varpropto\beta
R^{0.54}$ at large rigidities $R>4.5$ GV in this model is probably not in
contradiction with the observations of interstellar turbulence. At the same
time, the sharp change-over of the diffusion to the regime $D=const$ at
rigidities $R<4.5$ GV can not be naturally explained by the kinetic theory of
particle scattering in random magnetic fields. The \textit{ad hoc} elucidation
of the observed peak in the secondary-to-pramary ratios makes the model implausible.

The reacceleration model is favored as regards to its natural description of the
energy dependence of particle transport in the Galaxy with a close to
Kolmogorov spectrum of turbulence since the scaling $D_{res}\varpropto\beta
R^{0.3}$ at all rigidities is assumed in this case. The structure of the peak
in the secondary-to-primary ratios is reproduced in this model. The model with
reacceleration has no problems with the account for low anisotropy of cosmic
rays. However, the predicted weak energy dependence of secondary-to-primary
ratios at energies $\gtrsim20$ GeV/n is not in agreement with the high energy
HEAO-3 data by \citet{Bietal81} on Sub-Fe/Fe ratio, see Figure 6.

The source spectra for primary nuclei derived in all propagation models are close to
$R^{-2.3} - R^{-2.4}$
that is uncomfortably steep compared with the predictions of supernova shock
acceleration models \citep{Betal99,BandE99}. Strictly speaking, this refers to the limited
energy range up to about
$30$ GeV/n where statistically accurate data on secondary and primary nuclei are
available. Even if one does
not assume that a single power law source spectrum continues to much higher energies,
the soft depends
of diffusion on energy in the reacceleration model imply a steep source spectrum close
to $R^{-2.4}$ to fit
the observed spectrum of primaries close to $R^{-2.75}$. This problem may be not so
pressing for the
disk-halo diffusion model.

It is clear that measurements of secondary to primary ratios at higher energy are needed since it is here that the models begin to seriously diverge. Such future missions such as ACCESS would be very helpful in determining which pictures of galactic cosmic-ray propagation are feasible and which are not.

\textbf{Acknowledgments.} The authors are grateful to S. A. Stephens and R. A.  Streitmatter for
providing a database of cosmic ray composition and energy spectra. We thank
Toru Shibata for giving us updated Sub-Fe/Fe ratios. This work was supported
by NASA grant NAG5-7069.

\newpage
\begin{center}
\bf Figure Captions
\end{center}
\setcounter{figure}{0}
\figcaption [Diffusion_model.eps] {Simplified galactic model with a diffusive halo,  galactic matter disk is of infinitessimal thickness.}

\figcaption [Turb_model.eps] {Simplified galactic model with a halo wind; galactic matter disk is of infinitessimal thickness.}

\figcaption [Wind_model.eps] {Turbulent diffusion model, halo is diffusive with two types of diffusion.}

\figcaption [Reacceleration_model.eps] {Reacceleration model, halo is diffusive in momentum as well as space.}

\figcaption [B2C.eps]{ The least squares fit to observed B/C ratios.
in four propagation models: turbulent diffusion (dashed lines), wind (dotted
lines), reacceleration (dash-dotted lines), and the disk-halo diffusion model
with the diffusion coefficient given by equation (\ref{Diffcoef}) (solid lines).  $\Phi$ is the force field approximation solar modulation parameter. Data are from a
comprehensive compilation by \citet{SS98}. }

\figcaption[subfe2fe.eps]{The least squares fit to observed  Sub-Fe/Fe ratios.
in four propagation models: turbulent diffusion (dashed lines), wind (dotted
lines), reacceleration (dash-dotted lines), and the disk-halo diffusion model
with the diffusion coefficient given by equation (\ref{Diffcoef}) (solid lines).  $\Phi$ is the force field approximation solar modulation parameter. Data are from a
comprehensive compilation by \citet{SS98}}

\figcaption[c.eps]{The least squares fit to the C spectrum in the same
propagation models as indicated in Figures (\ref{B2C}) and (\ref{sFe2Fe}). }

\figcaption[fe.eps]{The least squares fit to the Fe spectrum in the same
propagation models as indicated in Figures (\ref{B2C}) and (\ref{sFe2Fe}). }

\end{document}